\documentclass[twocolumn,showpacs,groupedaddress,assymb,amsmath]{revtex4}
\usepackage{graphicx}
\usepackage{amssymb}
\usepackage{amsmath}
\usepackage[none]{hyphenat}
\usepackage{mathrsfs}

\begin{document}

\title{$n$-photon blockade with an $n$-photon parametric drive}
\author{Yan-Hui Zhou$^{1}$, Fabrizio Minganti$^{2}$, Wei Qin$^{2}$, Qi-Cheng Wu$^{1}$, Junlong Zhao$^{1}$, Yu-Liang Fang$^{1}$, Franco Nori$^{2,3}$\footnote{Corresponding  address: fnori@riken.jp}, and
Chui-Ping Yang$^{1,4}$\footnote{Corresponding  address: yangcp@hznu.edu.cn}}
\affiliation{\centerline{$^1$ Quantum Information Research Center, Shangrao Normal University, Shangrao 334001, China}\\
\centerline{$^2$ Theoretical Quantum Physics Laboratory, RIKEN Cluster for Pioneering Research, Wako-shi, Saitama 351-0198, Japan}\\
\centerline{$^3$ Physics Department, The University of Michigan, Ann Arbor, Michigan 48109-1040, USA}\\
\centerline{$^4$ Department of Physics, Hangzhou Normal University, Hangzhou 311121, China}\\
}

\date{\today}
\begin{abstract}
We propose a mechanism to engineer an $n$-photon blockade in a nonlinear cavity with an $n$-photon parametric
drive $\lambda(\hat{a}^{\dag n}+\hat{a}^n)$. When an $n$-photon-excitation resonance condition is satisfied, the presence of n photons
in the cavity suppresses the absorption of the subsequent photons. To confirm the validity of this proposal,
we study the n-photon blockade in an atom-cavity system, a Kerr-nonlinear resonator, and two-coupled Kerr nonlinear resonators. Our results demonstrate that $n$-photon bunching and $(n+1)$-photon antibunching can be
simultaneously obtained in these systems. This effect is due both to the anharmonic energy ladder and to the
nature of the $n$-photon drive. To show the importance of the drive, we compare the results of the $n$-photon drive
with a coherent (one-photon) drive, proving the enhancement of antibunching in the parametric-drive case. This
proposal is general and can be applied to realize the $n$-photon blockade in other nonlinear systems.
\end{abstract}

\pacs{42.50.Ar, 42.50.Pq}
 \maketitle



\section{Introduction}

In a nonlinear cavity driven by a classical
light field, the single-photon existence in the cavity blocks the creation of a second photon~\cite{r1,xu000,xu001}, which is known as the single-photon blockade (1PB). Due to its potential applications in information and communication technology, 1PB has been extensively studied in the past years~\cite{r21,r22,r23,r24,r25,r26,shi01,shi02,r812,t01}. For example, the PB has been
predicted in cavity quantum electrodynamics~\cite{n03,n04,A01},
quantum optomechanical system~\cite{n01,r30,rr00,guang1}, and second order nonlinear system~\cite{zhou01,zhou02,zhou03,zhou04,zhou05}.

Traditionally, realizing 1PB requires a \emph{large nonlinearity} to change
the energy-level structure of the system, and 1PB can be used to create a single-photon source~\cite{r31}.
The 1PB effect was first observed in an optical cavity coupled to a single trapped atom~\cite{r3}. Since then, many experimental groups have observed this strong antibunching behavior in different systems, including a photonic crystal~\cite{r4} and a superconducting circuit~\cite{r5}. In addition, the 1PB can also enable
by another mechanism, i.e., the quantum interference~\cite{h022,h023,h024,h241,h242,r94}, which has been recently observed experimentally~\cite{h025,h026}. In this paper, we are only concerned with the photon blockade based on energy level splitting due to the large nonlinearity.

The $n$-photon blockade ($n$PB) was proposed with the development of 1PB. In analogy to 1PB, $n$PB ($n\geq2$) is defined by the existence of $n$ photons in a nonlinear cavity suppressing the creation of subsequent photons. The $2$PB ($n$PB with $n=2$) was studied in a Kerr-type system driven
by a laser~\cite{r7}, in a strong-coupling qubit-cavity system~\cite{r71}, and in a cascaded cavity QED system~\cite{r11}. The 2PB can also be generated by squeezing~\cite{dc1}. Experimentally, 2PB was realized in an optical cavity strongly coupled to a single atom~\cite{r8}, where driving the atom gives a larger optical nonlinearity than driving the cavity. $n$PB with $n>2$ has been studied in a
cavity strongly coupled
to two atoms~\cite{r9},  in a cavity with two cascade three-level atoms~\cite{r10},  and in a Kerr-type system driven
by a laser~\cite{r811,r81}. Meanwhile, in analogy to photon blockades, the phonon blockades have been widely studied~\cite{sheng01,sheng02,sheng03,sheng04,sheng05}.

In this paper, we theoretically propose that $n$PB can be triggered in a
nonlinear cavity with $n$-photon parametric drive. For convenience, we denote ``$n$-photon parametric drive" as $n$PD. We first give a brief introduction to this proposal and then confirm its validity by considering three examples, i.e., an atom-cavity system, a single mode Kerr-nonlinearity system, and a two-coupled-cavities Kerr-nonlinearity system. This proposal is quite general and can be extended to other nonlinear systems for studying $n$PB via $n$PD. The study of the $n$PB in recent decades has mainly focused on a coherent (i.e., single-photon) driving.  Comparing with a proposal using a coherent driving, the use of a $n$PD has the following advantages: (i) The nonlinear systems like atom-cavity system will not exist $n$PB with a coherent driving to the cavity due to the
bosonic enhancement of photon~\cite{r8}, while we find that the $n$PB will exist in these system with a $n$PD, so the proposal with the $n$PD is more general to realize a $n$PB. (ii) In the same nonlinear system, the $n$PD approach has a stronger ($n+1$)-photon bunching than the coherent driving approach, so the $n$PD approach has a better $n$PB effect.

The remainder of this paper is organized as follows. In
Sec.~{\rm II}, we introduce the Proposal for $n$PB with $n$PD. In Sec.~{\rm
III}, we illustrate the $nPB$ in an atom-cavity system.
In Sec.~{\rm
\rm IV}, we show the $nPB$ in single-mode Kerr-nonlinearity system. In
Sec.~{\rm V}, we study the $2PB$ in two-coupled-cavities Kerr-nonlinearity system. Conclusion are given in
Sec.~{\rm VI}.

\section{Proposal for $n$PB with $n$PD}

The $n$PD with $n=2$ has many applications, such as in the realization of quantum metrology~\cite{r92} and cooling of a micromechanical mirror~\cite{r93}. In the following, we will present our basic idea for studing the $n$PB via $n$PD on a nonlinear cavity.

$n$PD involved in our proposal is described by $\hat{H}_d=\lambda(\hat{a}^{\dag n} e^{-i\omega_p t}+\hat{a}^ne^{i\omega_p t})$, where $\hat{a}$ is the cavity annihilation operator, $\lambda$ is the parametric driving amplitude, and $\omega_p$ is the driving frequency. Apart from the cavity on which $n$PD is applied, an auxiliary nonlinear system (e.g., an atom, a Kerr-nonlinearity medium, or an auxiliary cavity) is required to realize $n$PB. The Hamiltonian of the auxiliary nonlinear system and the cavity is denoted by $\hat{H}_0$. The form of $\hat{H}_0$ is not unique, and it depends on the type of the nonlinear system. Generally speaking, the Hamiltonian $\hat{H}_0$ can be diagonalized and expressed as
\begin{eqnarray}
\hat{H}_0&=&\sum_{j=1}^{k_1}\omega_1^j|\psi_1^j\rangle\langle\psi_1^j|+\sum_{j=1}^{k_2}\omega_2^j|\psi_2^j\rangle\langle\psi_2^j|+\nonumber\\
&&\cdots+\sum_{j=1}^{k_n}\omega_n^j|\psi_n^j\rangle\langle\psi_n^j|+\cdots,
\label{1}
\end{eqnarray}
where $\omega_n^j$ is the $j$th eigenfrequency of $\hat{H}_0$ for the photon excitation number $n$, and we have assumed that the ground state energy is zero. The corresponding eigenstate $|\psi_n^j\rangle$ is constructed by the $k_n$ basis for $n$-photon excitation, where the basis
forms a closed space. The set of eigenfrequencies $\{\omega_1^j\},\{\omega_2^j\}\cdots,\{\omega_n^j\},\cdots$ are anharmonic due to the strong nonlinear interaction.
Among these eigenfrequencies, \{$\omega_n^j$\} (where $j$ is from $1$ to $k_n$) is crucial to $n$PB because the corresponding eigenstate \{$|\psi\rangle_{n}^j$\} includes a $n$-photon state. When the parametric drive frequency $\omega_p$ is tuned to the \{$\omega_n^j$\}, the parametric drive resonantly excites $n$ photons in the cavity. As a result, the system occupies the state \{$|\psi\rangle_{n}^j$\} via the nonlinear interaction. This gives rise to an important result for $n$PB. The corresponding conditions for $n$PB are
\begin{eqnarray}
\omega_p=\omega_n^1,~~~
\omega_p=\omega_n^2,
~~~\cdots ~~~
\omega_p=\omega_n^{k_n},
\label{101}
\end{eqnarray}
The $n$-photon resonance excitation by $n$PD ensures that the $n$-photon blockade is triggered in the nonlinear cavity.

To verify the validity of the above proposal, we will study three examples to study $n$PB, in: an atom-cavity system, a single-mode Kerr-nonlinearity system, and a two-coupled-cavities Kerr-nonlinearity system. In these systems, the analytical conditions for $n$PB and the accurate numerical results are studied, which conform that $n$PB can be triggered in a nonlinear cavity with $n$PD if the Hamiltonian $\hat{H}_0$ can be diagonalized.

The numerical confirmation of $n$PB adopts an equal-time correlation function, the equal-time $n$-order correlation function is defined as $g^{(n)}(0)=\langle
\hat{a}^{\dag n}\hat{a}^n\rangle/\langle \hat{a}^{\dag}\hat{a}\rangle^n$. The correlation function is calculated by numerically solving the master equation
in the steady state.
In order to prove $n$PB, it is sufficient
to fulfill the conditions $g^{(n)}(0)\geq0$ and $g^{(n+1)}(0)<0$ simultaneously~\cite{r8}.

\section{Atom-cavity system}

The atom-cavity system is described by the Jaynes-Cummings Hamiltonian, where the cavity is  driven by a $n$PD. In a frame rotating at the
parametric drive frequency $\omega_p/n$, the Hamiltonian is (assuming $\hbar=1$ hereafter)
\begin{eqnarray}
\hat{H}=\Delta_a\hat{a}^{\dag}\hat{a}+\Delta_e\hat{\sigma}_+\hat{\sigma}_-
+ g(\hat{a}^{\dag}\hat{\sigma}_-+\hat{\sigma}_+\hat{a})
+ \lambda(\hat{a}^{\dag n}+\hat{a}^n),
\label{001}
\end{eqnarray}
where $\hat{a}$ is the cavity annihilation operator, $\hat{\sigma}_{\pm}$ are the atom
raising and lowering operators, $g$ is the coupling strength of the atom and the cavity mode, $\lambda$ is the amplitude of $n$PD, and $\Delta_a=\omega_a-\omega_p/n$ ($\Delta_e=\omega_e-\omega_p/n$) is the detuning between the cavity frequency $\omega_a$ (the atom frequency $\omega_e$) and the $1/n$ driving frequency. Here and below, we study the case of $\omega_a=\omega_e$ for convenience, resulting in $\Delta_a=\Delta_e$. The Hamiltonian (\ref{001}) with $n=2$ can be used to exponentially enhance the light-matter coupling in a generic cavity QED~\cite{p0,p1,p2}.

\begin{figure}
\centering
\includegraphics[scale=0.58]{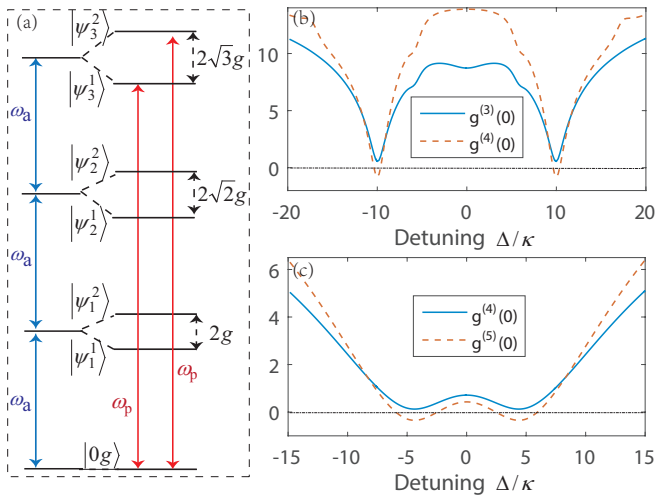}
\caption{(Color online)
(a) Schematic energy-level diagram explaining the occurrence of 3PB. (b) The logarithmic plot (of base e) of three-order correlation function $g^{(3)}(0)$ and fourth-order correlation function $g^{(4)}(0)$ as a function of detuning $\Delta/\kappa$, for $g/\kappa=10\sqrt{3}$, $\gamma/\kappa=0.1$, and $\lambda/\kappa=0.3$. (c) $g^{(4)}(0)$ and $g^{(5)}(0)$ as a function of $\Delta/\kappa$, for $g/\kappa=10$, $\gamma/\kappa=0.1$, and $\lambda/\kappa=1.5$.
 }
\label{fig1}
\end{figure}

In the absence of the $n$PD, the atom-cavity Hamiltonian $\hat{H}_0$ (the first three terms of Eq.~(\ref{001}) without driving) is diagonalized as
\begin{eqnarray}
\hat{H}_0&=&\sum_{j=1}^{2}\omega_1^j|\psi_1^j\rangle\langle\psi_1^j|+\sum_{j=1}^{2}\omega_2^j|\psi_2^j\rangle\langle\psi_2^j|+\nonumber\\
&&\cdots+\sum_{j=1}^{2}\omega_n^j|\psi_n^j\rangle\langle\psi_n^j|+\cdots.
\label{0010}
\end{eqnarray}
The energy
eigenstates of the system are $|\psi_n^{1,2}\rangle=1/\sqrt{2}(|n-1,e\rangle\mp|n,g\rangle)$, where $|g\rangle$ ($|e\rangle$) is the ground (excited) state of the atom, $n$ denotes the photon excitation number. For a $n$-photon excitation, the basis \{$|n,g\rangle$, $|n-1,e\rangle$\} forms a closed space. The corresponding eigenfrequencies with the $n$-photon excitation are $\omega_{n}^{1,2}=n\omega_a\mp\sqrt{n}g$.
The energy-level diagram of the system is shown in Fig.~\ref{fig1}(a).
The optimal conditions for $n$PB are calculated according to Eq.~(\ref{101}), which are simplified as
\begin{eqnarray}
g=\pm \sqrt{n}\Delta,
\label{002}
\end{eqnarray}
where $\Delta=\Delta_a=\Delta_e$.
There is one path for the system to reach the state $|\psi_n^{1,2}\rangle$: the system
first arrives at a $n$-photon state by $n$PD, then goes to
the state of $|\psi_n^{1,2}\rangle$ via the
coupling $g$, i.e., $|0g\rangle\stackrel{\lambda}{\longrightarrow}|ng\rangle\stackrel{g}{\longrightarrow}|\psi_n^{1,2}\rangle$, the $n$PD and the $n$-photon resonance excitation make that the $n$PB is triggered.

Next, we numerically study the $n$PB effect. The system dynamics is governed by the master equation $\partial\hat{\rho}/\partial t=-i[\hat{H},\hat{\rho}]+\kappa\ell(\hat{a})\rho+\gamma\ell(\hat{\sigma_-})\rho$,
where $\kappa$ denotes the decay rate of the cavity and $\gamma$ is the atomic spontaneous emission rate. The superoperators
are defined by $\ell(\hat{o})\hat{\rho}=\hat{o}\hat{\rho}\hat{o}^{\dag}-\frac{1}{2}\hat{o}^{\dag}\hat{o}\hat{\rho}-\frac{1}{2}\hat{\rho}\hat{o}^{\dag}\hat{o}$.
The numerical solutions of $g^{(n)}(0)$ and $g^{(n+1)}(0)$ are calculated by solving the master equation in the steady state. In Fig.~\ref{fig1}(b), we study a 3PB by plotting $g^{(3)}(0)$ and $g^{(4)}(0)$ versus $\Delta/\kappa$ with $g/\kappa=10\sqrt{3}$.
We note that the 3PB appears on $\Delta/\kappa=\pm10$ ($g^{(3)}(0)\geq0$ and $g^{(4)}(0)<0$ simultaneously), which agrees well with the conditions for $n$PB in Eq.~(\ref{002}) with $n=3$. The 4PB is studied in Fig.~\ref{fig1}(c) with $g/\kappa=10$, and 4PB appears on $\Delta/\kappa=\pm5$, which also agrees with Eq.~(\ref{002}) with $n=4$. The numerical results confirm the analytic conditions and the corresponding analysis. In the above atom-cavity system, it was proved that the $n$PB will not exist with a coherent driving (driving the cavity) due to a consequence of the bosonic enhancement of photon~\cite{r8}, while the $n$PB will exist for this system with a $n$PD. So the proposal with the $n$PD is more general and the $n$PB will occur as long as the analytical eigenvalues of the nonlinear Hamiltonian $\{\omega_n^j\}$ is solvable.

\section{Single-mode Kerr-nonlinearity system}

The system of a single-mode cavity with a Kerr nonlinearity, driven by $n$PD with $n=2$, has been extensively studied due to its rich physics~\cite {p4,p5, p6,p7,p8}. Here we investigate $n$PB utilizing this system.
The Hamiltonian of this model in a
rotating frame is written as~\cite{p5}
\begin{eqnarray}
\hat{H}=\Delta\hat{a}^\dag\hat{a}+U\hat{a}^\dag\hat{a}^\dag\hat{a}\hat{a}+\lambda(\hat{a}^{\dag n}+\hat{a}^n),
\label{003}
\end{eqnarray}
where $\Delta_a=\omega_a-\omega_p/n$ is the cavity
detuning from the $1/n$ driving eigenfrequency, $U$ is the Kerr nonlinear strength, and $\lambda$ is the amplitude of the $n$PD.

\begin{figure}
\centering
\includegraphics[scale=0.65]{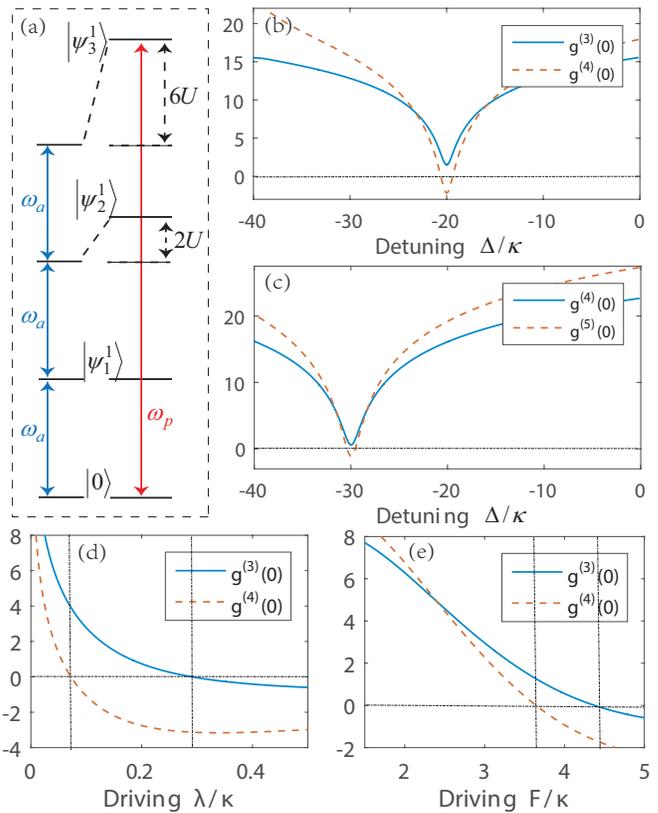}
\caption{(Color online)
(a) Energy spectrum of the single mode Kerr-nonlinearity system leading to 3PB via $3$PD.
(b) The logarithmic plot of $g^{(3)}(0)$ and $g^{(4)}(0)$ as a function of  $\Delta/\kappa$. (c) The logarithmic plot of $g^{(4)}(0)$ and $g^{(5)}(0)$ as a function of  $\Delta/\kappa$. In (b, c),
the parameters are $U/\kappa=10$ and $\lambda/\kappa=0.1$. (d) and (e) The logarithmic plot of $g^{(3)}(0)$ and $g^{(4)}(0)$ as a function of $\lambda/\kappa$ ($F/\kappa$) for $U/\kappa=10$ and $\Delta/\kappa=-20$.
 }
\label{fig2}
\end{figure}

The undriven part of the Hamiltonian (\ref{003}) is diagonalized as
\begin{eqnarray}
\hat{H}_0&=&\omega_1^1|\psi_1^1\rangle\langle\psi_1^1|+\omega_2^1|\psi_2^1\rangle\langle\psi_2^1|+\cdots\nonumber\\
&&+\omega_n^1|\psi_n^1\rangle\langle\psi_n^1|+\cdots,
\label{00301}
\end{eqnarray}
where the eigenstate is written as the Fock-state basis $|\psi_n^1\rangle=|n\rangle$ (with $n$ photons in the cavity), the corresponding eigenfrequency is $\omega_{n}^1=\omega_a n+U(n^2-n)$. The $n$PB can be triggered by the $n$-photon-excitation resonance, and the $|0\rangle\rightarrow|n\rangle$ transition is enhanced. The condition for $n$PB is obtained according to Eq.~(\ref{101}), which is given by
\begin{eqnarray}
U=-\frac{\Delta}{n-1}.
\label{004}
\end{eqnarray}
Because of the $n$PD and the $n$-photon-excitation resonance, the $n$ photon probability will increase when the condition (\ref{004}) is satisfied, and the $n$PB is triggered.

The master equation for the system is given by $\partial\hat{\rho}/\partial t=-i[\hat{H},\hat{\rho}]+\kappa\ell(\hat{a})\rho$. The energy-level diagram for 3PB is shown in Fig.~\ref{fig2}(a), and the corresponding numerical simulation is shown in Fig.~\ref{fig2}(b), where we plot $g^{(3)}(0)$ and $g^{(4)}(0)$ as a function of $\Delta/\kappa$ with $g/\kappa=10$. These results show that 3PB can be obtained at $\Delta/\kappa=-20$, as predicted in Eq.~(\ref{004}) for $n=3$.  The 4PB is studied in Fig.~\ref{fig2}(c) and the 4PB appears on $\Delta/\kappa=-30$, which also agrees with Eq.~(\ref{004}) with $n=4$.

We note that the studies to date on the $n$PB
are mainly focused on a coherent driving $F(\hat{a}^\dag+\hat{a})$, where $F$ is the coherent driving strength. So we compare the 3PB based on the $3$PD with that based on the coherent driving. To this end, we plot $g^{(3)}(0)$ and $g^{(4)}(0)$ versus the $3$PD strength and coherent driving strength in Fig.~\ref{fig2}(d, e) under the blockade condition of Eq.~(\ref{004})  ($g/\kappa=10$, $\Delta/\kappa=-20$), respectively.
The 3PB with the $3$PD is obtained in a region of small $\lambda$, while the implementation of 3PB with coherent driving needs a larger $F$. The most
striking feature is that the 3PB with the $3$PD
has a stronger four-photon antibunching and
three-photon bunching.


\section{Two-coupled-cavities Kerr-nonlinearity system}

Two coupled cavities with Kerr nonlinearity were considered to study 1PB~\cite{p9}.  We define the two cavities as cavities $a$ and $b$. The Hamiltonian in a rotating frame is
\begin{eqnarray}
\hat{H}&=&\Delta\hat{a}^\dag\hat{a}+\Delta\hat{b}^\dag\hat{b}+J(\hat{a}^\dag\hat{b}+\hat{b}^\dag\hat{a})+U(\hat{a}^\dag\hat{a}^\dag\hat{a}\hat{a}+\hat{b}^\dag\hat{b}^\dag\hat{b}\hat{b})\nonumber\\
&&+\lambda(\hat{a}^{\dag n}+\hat{a}^n),
\label{005}
\end{eqnarray}
where $\hat{a}$ ($\hat{b}$) is the photon annihilation operator for cavity $a$ ($b$) with frequency $\omega_a$ ($\omega_b$), $\Delta=\omega_a-\omega_p/n=\omega_b-\omega_p/n$, $J$ is the coupling strength of the two cavities, $U$ is the Kerr nonlinear strength, and $\lambda$ is the $n$PD strength.

\begin{figure}
\centering
\includegraphics[scale=0.56]{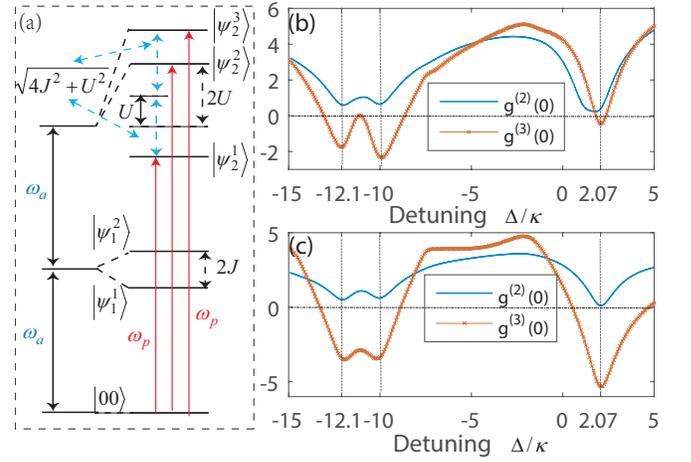}
\caption{
(a) Energy spectrum for two coupled cavities with Kerr nonlinearity. (b, c) The logarithmic plot (of base e) of $g^{(2)}(0)$ and $g^{(3)}(0)$ as a function of $\Delta/\kappa$ for cavity $a$ and cavity $b$, respectively. (b) Cavity $a$. (c) Cavity $b$. In (b, c), the parameters are $U/\kappa=10$, $J/\kappa=5$, and $\lambda/\kappa=0.5$.
 }
\label{fig3}
\end{figure}

The Hamiltonian for the two cavities with the Kerr nonlinearity (the first four terms in Eq.~(\ref{005}) without driving) is diagonalized as
\begin{eqnarray}
\hat{H}_0&=&\sum_{j=1}^{2}\omega_1^j|\psi_1^j\rangle\langle\psi_1^j|+\sum_{j=1}^{3}\omega_2^j|\psi_2^j\rangle\langle\psi_2^j|+\nonumber\\
&&\cdots+\sum_{j=1}^{n+1}\omega_n^j|\psi_n^j\rangle\langle\psi_n^j|+\cdots.
\label{0010}
\end{eqnarray}
We find that our approach comes with its own limitations in this system. The eigenfrequencies $\{\omega_n^j\}$ are more and more difficult to analytically solve with the increase of $n$, so we only study the case of $n=2$, the corresponding energy-level diagram is shown in Fig.~\ref{fig3}(a).
Now we derive the eigenfrequencies $\{\omega_2^j\}$ and the eigenstates $\{|\psi_2^j\rangle\}$.
To obtain these, the Hamiltonian will be expanded with the two-cavity states $|20\rangle$, $|02\rangle$ and $|11\rangle$ for the two-photon excitation, where $|\alpha\beta\rangle$ is the Fock-state basis of the system with the number $\alpha$ ($\beta$) denoting the photon number in cavity $a$ ($b$). The two-cavity states satisfy the two-photon excitation condition $\alpha+\beta=2$, and the states $|20\rangle$, $|02\rangle$ and $|11\rangle$ form a closed space. Under these basis states, the Hamiltonian with two-photon excitation can be described as
\begin{eqnarray}
\hat{H}_2=
\begin{bmatrix}
2\omega_a+2U & \sqrt{2}J & 0\\
\sqrt{2}J & 2\omega_a & \sqrt{2}J \\
0         & \sqrt{2}J        & 2\omega_a+2U
\end{bmatrix}. \label{06}
\end{eqnarray}
The three eigenfrequencies are $\omega_2^2=2(U+\omega_a)$, and $\omega_{2}^{1,3}=2\omega_a+U\mp\sqrt{4J^2+U^2}$. The corresponding unnormalized eigenstates are $|\psi_2^2\rangle=-|20\rangle+|02\rangle$, and $|\psi_2^{1,3}\rangle=|20\rangle-[\sqrt{2}U\mp\sqrt{2(4J^2+U^2)}]/(2J)|11\rangle+|02\rangle$.
The conditions for 2PB, obtained from Eq.~(\ref{101}), are given by
\begin{eqnarray}
\Delta=-U,~~~~~
\Delta=\frac{-U\pm\sqrt{4J^2+U^2}}{2}.
\label{007}
\end{eqnarray}
Under these resonance conditions, 2PB can be triggered,
which enhances the transition $|00\rangle\rightarrow\{|\psi_2^2\rangle, |\psi_2^{1,3}\rangle\}$.
The two cavities occupy the two-photon states $|20\rangle$ and $|02\rangle$, which ensures that 2PB is simultaneously realized in the two cavities when the conditions (\ref{007}) are satisfied.

The numerical study of 2PB is the same as before. In Fig.~\ref{fig3}(b, c), we plot $g^{(2)}(0)$ and $g^{(3)}(0)$ as a function of $\Delta/\kappa$ for cavity $a$ and cavity $b$, respectively. The results indicate that 2PB occurs for $\Delta/\kappa=-12.7$, $\Delta/\kappa=-10$ and $\Delta/\kappa=2.07$, which are predicted by the three $n$PB conditions given in Eq.~(\ref{007}) with $n=2$.
The anharmonic distribution of the blockade points are determined by the anharmonic splitting of the energy levels $\omega_2^1$, $\omega_2^2$, and $\omega_2^3$. The distance of the two blockade points on the left is calculated as $d=\sqrt{4J^2+U^2}-U$, and the distance of the two points on the right is $d=\sqrt{4J^2+U^2}+U$. Thus, it can be concluded that 2PB is simultaneously realized in both cavity $a$ and cavity $b$ due to the feature of the system and the $N$PD.

\begin{figure}
\centering
\includegraphics[scale=0.79]{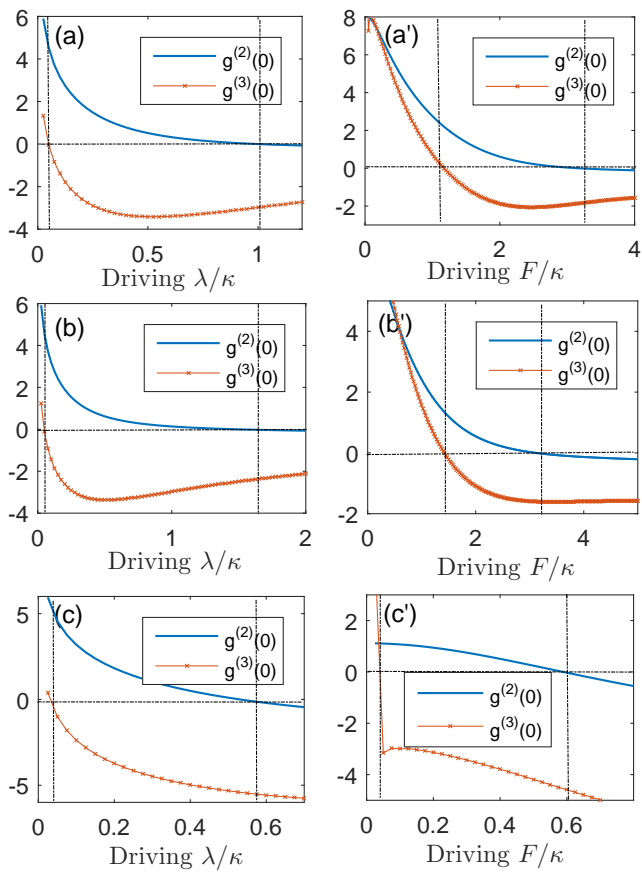}
\caption{ The logarithmic plot of $g^{(2)}(0)$ and $g^{(3)}(0)$ of cavity $b$ as a function of $\lambda/\kappa$ ($F/\kappa$) for $U/\kappa=10$ and $J/\kappa=5$.
(a, a') $\Delta/\kappa=-12.5$. (b, b') $\Delta/\kappa=-10$. (c, c') $\Delta/\kappa=2.07$.
 }
\label{fig4}
\end{figure}

The undriven cavity $b$ has a better 2PB effect than cavity $a$ for a smaller $g^{(3)}(0)$ shown in Fig.~\ref{fig3}(b, c), so we compare the $2$PD approach with the coherent driving approach for cavity $b$. The results are presented in Fig.~\ref{fig4}, where we plot of $g^{(2)}(0)$ and $g^{(3)}(0)$ as a function of $\lambda/\kappa$ ($F/\kappa$) under the three blockade conditions, respectively. We find that the two approaches have different blockade regions. And the same conclusion is arrived as the single-mode Kerr-nonlinearity system that the 2PB with the $2$PD has a stronger three-photon antibunching and two-photon bunching.

\section{Conclusion}

We have proposed that $n$-photon blockade can be realized
in a nonlinear cavity with a $n$-photon parametric drive. The validity of this proposal is confirmed by three examples, i.e., $n$-photon blockade in an atom-cavity system, in a single-mode Kerr nonlinear
device, and in a two-coupled-cavities Kerr-nonlinear system. By solving the master equation in the
steady-state limit and computing the correlation functions $g^{(n)}(0)$ and $g^{(n+1)}(0)$, we have shown that $n$PB can be realized, and
the optimal conditions for $n$PB are in good agreement
with the numerical simulations, which clearly illustrates the validity of our proposal.
This proposal can be extended
to other nonlinear systems, as long as the $n$-photon-excitation analytical eigenvalues of the nonlinear
Hamiltonian is solvable.
\\

This work is supported by the Key R\&D Program of Guangdong province (Grant No. 2018B0303326001), the NKRDP of china (Grants Number 2016YFA0301802), the National Natural Science Foundation of China (NSFC) under Grants No. 11965017, 11705025,11804228, 11774076, the Jiangxi Natural Science Foundation under Grant No. 20192ACBL20051, the Jiangxi Education Department Fund under Grant No. GJJ180873. This work is also supported by the NTT Research, Army Research Office (ARO) (Grant No. W911NF-18-1-0358), the Japan Science and Technology Agency (JST) (via the CREST Grant No. JPMJCR1676), the Japan Society for the Promotion of Science (JSPS) (via the KAKENHI Grant Number JP20H00134, JSPS-RFBR Grant No. 17-52-50023), the Grant No. FQXi-IAF19-06 from the Foundational Questions Institute Fund (FQXi), and a donor advised fund of the Silicon Valley Community Foundation.


\begin{references}

\bibitem{r1}A. Imamo\={g}lu, H. Schmidt, G. Woods, and M. Deutsch, Strongly interacting photons in a nonlinear cavity, Phys. Rev. Lett. \textbf{79}, 1467 (1997).

\bibitem{xu000}X.-W. Xu and Y. Li, Strong photon antibunching of symmetric and antisymmetric modes in weakly nonlinear photonic molecules, Phys. Rev. A \textbf{90}, 033809 (2014).
\bibitem{xu001}O. Kyriienko, I. A. Shelykh, and T. C. H. Liew, Tunable single-photon emission from dipolaritons,
Phys. Rev. A \textbf{90}, 033807 (2014).


\bibitem{r21}H. J. Kimble, M. Dagenais, and L. Mandel, Photon antibunching in resonance fluorescence, Phys. Rev. Lett.
\textbf{39}, 691 (1977).

\bibitem{r22}B. Dayan, A. S. Parkins, T. Aoki, E. P. Ostby, K. J. Vahala,
and H. J. Kimble, A photon turnstile dynamically regulated by one atom, Science \textbf{319}, 1062 (2008).


\bibitem{r23}A. Reinhard, T. Volz, M. Winger, A. Badolato, K. J.
Hennessy, E. L. Hu, and A. Imamoglu, Strongly correlated photons on a chip, Nat. Photonics \textbf{6}, 93 (2011).
\bibitem{r24} C. Lang, D. Bozyigit, C. Eichler, L. Steffen, J. M. Fink,
A. A. Abdumalikov, M. Baur, S. Filipp, M. P. da Silva, A.
Blais, and A. Wallraff, Observation of resonant photon blockade at microwave frequencies using correlation function measurements, Phys. Rev. Lett. \textbf{106}, 243601 (2011).

\bibitem{r25}H. J. Carmichael, Breakdown of photon blockade: A dissipative quantum phase transition in zero dimensions, Phys. Rev. X \textbf{5}, 031028 (2015).

\bibitem{r26}C. Vaneph, A. Morvan, G. Aiello, M. F\'{e}chant, M. Aprili, J. Gabelli, and J. Est\`{e}ve, Observation of the Unconventional Photon Blockade in the Microwave Domain, Phys. Rev. Lett. \textbf{121}, 043602 (2018).






\bibitem{shi01}S. L. Su, Y. Z. Tian, H. Z. Shen, H. P. Zang, E. J. Liang, and S. Zhang, Applications of the modified Rydberg antiblockade regime with simultaneous driving, Phys. Rev. A \textbf{96}, 042335 (2017).
\bibitem{shi02}  S. L. Su, Y. Gao, E. J. Liang, and S. Zhang, Fast Rydberg antiblockade regime and its applications in quantum logic gates, Phys. Rev. A \textbf{95}, 022319 (2017).
\bibitem{r812}A. Miranowicz, J. Bajer, M. Paprzycka, Y.X. Liu, A.M. Zagoskin, F. Nori, State-dependent photon blockade via quantum-reservoir engineering, Phys. Rev. A \textbf{90}, 033831 (2014).
\bibitem{t01}Y.X. Liu, X.W. Xu, A. Miranowicz, F. Nori, From blockade to transparency: Controllable photon transmission through a circuit-QED system, Phys. Rev. A \textbf{89}, 043818 (2014).



\bibitem{n03}Y. X. Liu, A. Miranowicz, Y. B. Gao, J. Bajer, C. P. Sun, F. Nori, Qubit-induced phonon blockade as a signature of quantum behavior in nanomechanical resonators, Phys. Rev. A \textbf{82}, 032101 (2010).
\bibitem{n04}Y. H. Zhou, H. Z. Shen, X. Y. Zhang, and X. X. Yi, Zero eigenvalues of a photon blockade induced by a non-Hermitian Hamiltonian with a gain cavity, Phys. Rev. A \textbf{97}, 043819 (2018)
\bibitem{A01}Y. X. Liu, A. Miranowicz, Y. B. Gao, J. Bajer, C. P. Sun, F. Nori, Qubit-induced phonon blockade as a signature of quantum behavior in nanomechanical resonators, Phys. Rev. A \textbf{82}, 032101 (2010).


\bibitem{n01}A. Nunnenkamp, K. B{\o}rkje, and S. M. Girvin, Single-Photon Optomechanics, Phys. Rev. Lett. \textbf{107}, 063602 (2011).
\bibitem{r30}P. Rabl, Photon Blockade Effect in Optomechanical Systems, Phys. Rev. Lett. \textbf{107}, 063601 (2011).

\bibitem{rr00}H. Wang, X. Gu, Y.X. Liu, A. Miranowicz, F. Nori, Tunable photon blockade in a hybrid system consisting of an optomechanical device coupled to a two-level system, Phys. Rev. A \textbf{92}, 033806 (2015).

\bibitem{guang1}J.Q. Liao, F. Nori, Photon blockade in quadratically coupled optomechanical systems, Phys. Rev. A \textbf{88}, 023853 (2013).










\bibitem{zhou01}Y. H. Zhou, H. Z. Shen, and X. X. Yi, Unconventional photon blockade with second-order nonlinearity,
Phys. Rev. A \textbf{92}, 023838 (2015).
\bibitem{zhou02}Y. H. Zhou, X. Y. Zhang, Q. C. Wu, B. L. Ye, Zhi-Qiang Zhang, D. D. Zou, H. Z. Shen, and C.-P. Yang,
Conventional photon blockade with a three-wave mixing, Phys. Rev. A \textbf{102}, 033713 (2020).
\bibitem{zhou03}H. Z. Shen, Y. H. Zhou, and X. X. Yi, Quantum optical diode with semiconductor microcavities, Phys. Rev. A \textbf{90}, 023849 (2014).
\bibitem{zhou04} Y. H. Zhou, H. Z. Shen, X. Q. Shao, and X. X. Yi, Opt. Express \textbf{24}, 17332 (2016).

\bibitem{zhou05}A. Majumdar and D. Gerace, Single-photon blockade in doubly resonant nanocavities with second-order nonlinearity, Phys. Rev. B \textbf{87}, 235319 (2013).




\bibitem{r31}A. Ridolfo, M. Leib, S. Savasta, and M. J. Hartmann, Photon Blockade in the Ultrastrong Coupling Regime,
Phys. Rev. Lett. \textbf{109}, 193602 (2012).

\bibitem{r3}K. M. Birnbaum, A. Boca, R. Miller, A. D. Boozer, T. E.
Northup, and H. J. Kimble, Photon blockade in an optical cavity with one trapped atom, Nature (London) \textbf{436}, 87 (2005).

\bibitem{r4}A. Faraon, I. Fushman, D. Englund, N. Stoltz, P. Petroff, and J.
Vu\u{c}kovi\`{c}, Coherent generation of non-classical light on a chip via photon-induced tunnelling and blockade, Nat. Phys. \textbf{4}, 859 (2008).

\bibitem{r5}A. J. Hoffman, S. J. Srinivasan, S. Schmidt, L. Spietz, J. Aumentado, H. E. Tureci, and A. A. Houck, Dispersive photon blockade in a superconducting circuit, Phys. Rev. Lett. \textbf{107}, 053602 (2011).
\bibitem{h022}T. C. H. Liew and V. Savona, Single Photons from Coupled Quantum Modes, Phys. Rev. Lett. \textbf{104}, 183601 (2010).
\bibitem{h023}H. J. Carmichael, Photon antibunching and squeezing for a single atom in a resonant cavity, Phys. Rev. Lett. \textbf{55}, 2790 (1985).

\bibitem{h024} M. Bamba, A. Imamoglu, I. Carusotto, and C. Ciuti, Origin of strong photon antibunching in weakly nonlinear photonic molecules, Phys. Rev. A \textbf{83}, 021802(R) (2011).

\bibitem{h241}H. Flayac and V. Savona, Input-output theory of the unconventional photon blockade, Phys. Rev. A \textbf{88}, 033836 (2013).

\bibitem{h242}H. Z. Shen, C. Shang, Y. H. Zhou, and X. X. Yi, Unconventional single-photon blockade in non-Markovian systems, Phys. Rev. A \textbf{98}, 023856 (2018).

 \bibitem{r94} B. Sarma and A. K. Sarma, Quantum-interference-assisted photon blockade in a cavity via parametric interactions, Phys. Rev. A \textbf{96}, 053827 (2017).

\bibitem{h025}H. J. Snijders, J. A. Frey, J. Norman, H. Flayac, V. Savona, A. C. Gossard, J. E. Bowers, M. P. van Exter, D. Bouwmeester, and W. L\"{o}ffler, Observation of the Unconventional Photon Blockade, Phys. Rev. Lett. \textbf{121}, 043601 (2018).

\bibitem{h026} C. Vaneph, A. Morvan, G. Aiello, M. F\'{e}chant, M. Aprili, J. Gabelli, and J. Est\`{e}ve, Observation of the unconventional photon blockade in the microwave domain, Phys. Rev. Lett. \textbf{121}, 043602 (2018).


\bibitem{r7}A. Miranowicz, M. Paprzycka, Y. Liu, J. Bajer, and F. Nori, Two-photon and three-photon blockades in driven nonlinear systems, Phys. Rev. A \textbf{87}, 023809 (2013).
\bibitem{r71}W.-W. Deng, G.-X. Li, and H. Qin, Enhancement of the two-photon blockade in a strong-coupling qubit-cavity system, Phys. Rev. A \textbf{91}, 043831 (2015).

\bibitem{r11}Q. Bin, X. L\"{u}, S.-W. Bin, and Y. Wu, Two-photon blockade in a cascaded cavity-quantum-electrodynamics system, Phys. Rev. A \textbf{98}, 043858 (2018).
\bibitem{dc1}A. Kowalewska-Kud{\l}aszyk, S.I. Abo, G. Chimczak, J. Pe\v{r}ina Jr., F. Nori, A. Miranowicz,
Two-photon blockade and photon-induced tunneling generated by squeezing,
Phys. Rev. A \textbf{100}, 053857 (2019).


\bibitem{r8}C. Hamsen, K. N. Tolazzi, T. Wilk, and G. Rempe, Two-photon blockade in an atom-driven cavity QED system, Phys. Rev.
Lett. \textbf{118}, 133604 (2017).



\bibitem{r9}C. J. Zhu, Y. P. Yang, and G. S. Agarwal, Collective multiphoton blockade in cavity quantum electrodynamics, Phys. Rev. A \textbf{95}, 063842 (2017).



\bibitem{r10}J. Z. Lin, K. Hou,1,3 C. J. Zhu, and Y. P. Yang, Manipulation and improvement of multiphoton blockade in a cavity-QED system with two cascade three-level atoms, Phys. Rev. A \textbf{99}, 053850 (2019).


\bibitem{r811}G. H. Hovsepyan, A. R. Shahinyan, and G. Yu. Kryuchkyan, Multiphoton blockades in pulsed regimes beyond stationary limits, Phys. Rev. A \textbf{90}, 013839 (2014).


\bibitem{r81}R. Huang, A. Miranowicz, J.-Q. Liao, F. Nori, and H. Jing, Nonreciprocal photon blockade, Phys. Rev. Lett. \textbf{121}, 153601 (2018).
\bibitem{sheng01}Y.X. Liu, A. Miranowicz, Y.B. Gao, J. Bajer, C.P. Sun, F. Nori, Qubit-induced phonon blockade as a signature of quantum behavior in nanomechanical resonators, Phys. Rev. A \textbf{82}, 032101 (2010).

\bibitem{sheng02}A. Miranowicz, J. Bajer, N. Lambert, Y.X. Liu, F. Nori, Tunable multiphonon blockade in coupled nanomechanical resonators, Phys. Rev. A \textbf{93}, 013808 (2016).
\bibitem{sheng03} X. Wang, A. Miranowicz, H.R. Li, F. Nori,
Method for observing robust and tunable phonon blockade in a nanomechanical resonator coupled to a charge qubit,
Phys. Rev. A \textbf{93}, 063861 (2016).

\bibitem{sheng04}Y.X. Liu, A. Miranowicz, Y.B. Gao, J. Bajer, C.P. Sun, F. Nori,
Qubit-induced phonon blockade as a signature of quantum behavior in nanomechanical resonators, Phys. Rev. A \textbf{82}, 032101 (2010).

\bibitem{sheng05} X. Wang, A. Miranowicz, H.R. Li, F. Nori,
Method for observing robust and tunable phonon blockade in a
nanomechanical resonator coupled to a charge qubit, Phys. Rev. A \textbf{93}, 063861 (2016).


\bibitem{r92} F. Hudelist, J. Kong, C. Liu, J. Jing, Z. Y. Ou, and W. Zhang,
Quantum metrology with parametric amplifier-based
photon correlation interferometers, Nat. Commun. \textbf{5}, 3049
(2014).

\bibitem{r93} S. Huang and G. S. Agarwal, Enhancement of cavity cooling of a micromechanical mirror using parametric interactions, Phys. Rev. A \textbf{79}, 013821 (2009).


\bibitem{p0}W. Qin, A. Miranowicz, P.-B. Li, X.-Y. L\"{u}, J. Q. You, and F. Nori, Exponentially Enhanced Light-Matter Interaction, Cooperativities, and Steady-State Entanglement Using Parametric Amplification, Phys. Rev. Lett. \textbf{120}, 093601 (2018).
\bibitem{p1} C. Leroux, L. C. G. Govia, and A. A. Clerk, Enhancing cavity quantum electrodynamics via antisqueezing: Synthetic ultrastrong coupling, Phys. Rev. Lett. \textbf{120}, 093602 (2018).



\bibitem{p2}Y. Wang, C. Li, E. M. Sampuli, J. Song, Y. Jiang, Y. Xia, Enhancement of coherent dipole coupling between two atoms via squeezing a cavity mode, Phys. Rev. A \textbf{99}, 023833 (2019).


\bibitem{p4}B. Wielinga and G. J. Milburn, Quantum tunneling in a Kerr medium with parametric pumping, Phys. Rev. A \textbf{48}, 2494 (1993).
\bibitem{p5}P. Zhao, Z. Jin, P. Xu, X. Tan, H. Yu, and Y. Yu, Two-photon driven Kerr resonator for quantum annealing with three-dimensional circuit QED, Phys. Rev. Appl. \textbf{10}, 024019 (2018).


\bibitem{p6} T. V. Gevorgyan and G. Y. Kryuchkyan, Parametrically driven nonlinear oscillator at a few-photon level, J. Mod.
Opt. \textbf{60}, 860 (2013).

\bibitem{p7}G. H. Hovsepyan, A. R. Shahinyan, L. Y. Chew, and G. Y. Kryuchkyan, Phase locking and quantum statistics in a parametrically driven nonlinear resonator, Phys. Rev. A \textbf{93}, 043856 (2016).

\bibitem{p8} N. Bartolo, F. Minganti, W. Casteels, and C. Ciuti, Exact steady state of a Kerr resonator with one-and two-photon driving and dissipation: Controllable Wigner-function multimodality and dissipative phase transitions, Phys. Rev. A \textbf{94}, 033841 (2016).

\bibitem{p9}M. Bamba, A. Imamoglu,  I. Carusotto, and C. Ciuti, Origin of strong photon antibunching in weakly nonlinear photonic molecules, Phys. Rev. A \textbf{83}, 021802(R) (2011).


\end{references}
\end{document}